\documentstyle[12pt]{article}

\begin{document}

\centerline{{\Large\bf The symmetry structure of the}}

\vspace{.2in}
\centerline{{\Large\bf anti-self-dual Einstein hierarchy}}

\vspace{.2in}
\centerline{{\bf I.A.B.Strachan}}

\vspace{.1in}
\centerline{Dept. of Mathematics and Statistics, University of Newcastle,}
\vspace{.1in}
\centerline{Newcastle-upon-Tyne, NE1 7RU, England
\footnote{e-mail: i.a.b.strachan@newcastle.ac.uk}.}

\vspace{.3in}
\centerline{{\bf Abstract}}

\vspace{.2in}
\small
\parbox{4.5in}{An important example of a multi-dimensional integrable
system is the
anti-self-dual Einstein equations. By studying the symmetries of these
equations, a recursion operator is found and the associated
hierarchy constructed. Owing to the properties of the recursion
operator one may construct a hierarchy of symmetries and find the
algebra generated by them. In addition, the Lax pair for this
hierarchy is constructed.}
\normalsize

\bigskip

\bigskip

\section*{1. Introduction }

\bigskip

One fundamental idea within the theory of integrable systems is that of a
hierarchy, the paradigm being the KdV hierarchy

\begin{equation}
u_t = {\cal R}^n u_x \,,\label{eq:kdv}
\end{equation}

\noindent where $\cal R$ is the recursion operator, defined by
${\cal R} = \partial_x^2 + 4 u + 2 u_x \partial_x^{-1} \,.$
One of the ways of introducing this operator is via the symmetries of the
original KdV equation, the quantities ${\cal R}^n u_x$ being generalized
symmetries of the KdV equation. Further analysis of the recursion operator
leads one to the concepts of commuting flows and the bi-Hamiltonian
structure of the KdV hierarchy.

\par

This structure persists in higher dimensional integrable systems. The
hierarchy associated with the anti-self-dual Yang-Mills equations has been
studied by many authors \cite{Ward.a,Strachan,Ablowitz.et.al,Schiff},
and has all the properties one would expect from an
integrable system -- the existence of recursion operators and bi-Hamiltonian
structure to name but two.

\par

The purpose of this paper is to study the hierarchical structure of
another important example of a multidimensional integrable system,
namely the anti-self-dual Einstein equations.
These define a 4-dimensional (possibly complex) metric of signature $(+,+,+,+)$
with the Ricci tensor zero and the Weyl tensor anti-self-dual \cite{Penrose}.
Such metrics have
an existence independent of the particular coordinate system used to describe
it, and so one of the problems in finding a anti-self-dual Einstein hierarchy
is
to find a suitable coordinate system in which the field equations take a
form similar to equation (\ref{eq:kdv}).
For example, all such metrics are automatically
K\"ahler \cite{Atiyah.et.al}, and so the metric may be written in terms of
a single scalar
function, the K\"ahler potential. The field equation is then

\begin{equation}
\Omega_{,wy}\Omega_{,xz} - \Omega_{,wz}\Omega_{,xy} = 1 \label{eq:plebanski}
\end{equation}

\noindent (and is known as Pleba\~nski's first heavenly equation
\cite{Plebanski}) and in this form it is
not at all obvious what the recursion operator should be, nor how to
construct the associated hierarchy.

\par

One possible way forward is via the use of the infinite dimensional Lie algebra
$sdiff(\Sigma^2)$ of area-preserving diffeomorphism of a 2-surface
$\Sigma^2\,,$
where one replaces the Lie bracket $[A,B] = AB - BA$ of matrices by the
Poisson bracket of functions

\begin{equation}
\{A,B\} = {\partial A\over\partial y} {\partial B\over\partial z} -
{\partial A\over\partial z}{\partial B\over\partial y}\,.   \label{eq:poisson}
\end{equation}

\noindent Park \cite{Park} has shown that one may
interpret~(\ref{eq:plebanski}) as a
2-dimensional topological chiral model with fields taking values in the
Lie algebra $sdiff(\Sigma^2)\,.$ As this chiral model is a reduction of the
anti-self-dual Yang-Mills equations one might hope that by studying an
appropriately
reduced version of the hierarchy one could obtain a chiral field hierarchy,
and then, via the replacement $[A,B]\rightarrow\{A,B\}$ to a anti-self-dual
Einstein
hierarchy. This convoluted approach will be avoided for two reason.
Firstly, underlying
these anti-self-dual integrable systems are two quite different geometric
constructions. Solutions to the anti-self-dual Yang-Mills equations are
encoded in
certain holomorphic vector bundles over a flat twistor space \cite{Ward.b},
while solutions
to the anti-self-dual Einstein equations are encoded within a curved twistor
space \cite{Penrose}.
Thus replacing matrix commutators by Poisson brackets only works at the level
of the differential equations, not in the underlying geometry. Secondly,
implicit in much of the work on anti-self-dual Yang-Mills hierarchies is the
fact
that the zero-curvature relation

\[
\partial_x A - \partial_y B + [A,B] = 0
\]

\noindent may be solved by taking
$A=g^{-1}\partial_y g\,,B=g^{-1}\partial_x g\,,$ for some group-valued element
$g\,.$ Unfortunately this does not hold under the replacement
$[A,B]\rightarrow\{A,B\}$, there being no notion of pure gauge.

\section*{2. The Cauchy-Kovalevski form of the anti-self-dual Einstein
equation}

By defining a new independent variable $t=-\Omega_w$ (so that now
$w=w(t,x,y,z)\,$)
and performing a Legendre transformation \cite{Finley.et.al} with

\[
h(t,x,y,z) = \Omega(w(t,x,y,z),x,y,z) + t\, w(t,x,y,z)\,,
\]

\noindent Pleba\~nski's first heavenly equation (\ref{eq:plebanski})
takes the new form

\begin{equation}
h_{tt} = h_{xz} h_{ty} - h_{xy} h_{tz}\,. \label{eq:granteqn}
\end{equation}

\medskip

\noindent This is in the Cauchy-Kovalevski form, and a formal solution may
be obtained as a power series in $t\,,$ depending on the initial data
$h|_{t=0}$ and $\partial_t h |_{t=0}\,.$ Equation (\ref{eq:granteqn}) was first
derived by Grant \cite{Grant}, using a different approach which will be given
in the next
section. The corresponding anti-self-dual Einstein metric is

\[
g=dt(h_{ty}dy + h_{tz}dz) + dx(h_{xy}dy+ h_{xz}dz) + {1\over h_{tt}}
(h_{ty}dy+h_{tz}dz)^2\,.
\]

To put equation (\ref{eq:granteqn}) into a form similar to (\ref{eq:kdv})
let $u=h_t\,,$ so $h_x=\partial^{-1}_t u_x\,.$ With this (\ref{eq:granteqn})
becomes

\begin{eqnarray*}
u_t & = & u_y (\partial^{-1}_t u_x)_z - u_z (\partial_t^{-1} u_x)_y \,, \\
    & = & [ (u_y \partial_z - u_z \partial_y ) \partial_t^{-1} ] u_x \,,
\end{eqnarray*}

\noindent which, with the help of (\ref{eq:poisson}), becomes

\begin{equation}
u_t = \{ u , \partial_t^{-1} u_x \}\,.  \label{eq:sdein}
\end{equation}

\medskip

\noindent This suggest that $\{u,\partial_t^{-1} ~~\}$ might be
a recursion operator. This is indeed the case, as will be shown in the next
two sections, and hence the anti-self-dual Einstein hierarchy is given by

\begin{equation}
u_t = \{ u , \partial_t^{-1} ~~ \}^n u_x\,.  \label{eq:sdeinhier}
\end{equation}

\medskip

Simple solutions to (\ref{eq:sdeinhier}) may be found quite easily. With
the ansatz

\[
u=-z + {\partial^n g(t,x,y)\over\partial t^n}
\]

\noindent the nonlinearities disappear, and one is left with a linear
equation for the function $g(t,x,y)\,:$

\[
{\partial^{n+1} g \over \partial t^{n+1}} -
{\partial^{n+1} g \over \partial y^n \partial x} = 0\,.
\]

\noindent Solutions to this equation may written in terms of a contour integral

\[
g(t,x,y) = \frac{1}{2\pi i} \oint k(\xi,y+\xi t + \xi^{n+1} x) d\xi\,,
\]

\noindent where $k(\xi,\omega)$ is an arbitrary holomorphic function.
This extends Bateman's formula for the $n=1$ case, which gives
metrics with anti-self-dual Killing vectors \cite{TodWard}.

\par

The appearance of the operator $\partial_t^{-1}$ in (\ref{eq:sdein}) is
somewhat unusual. On interchanging $x$ and $t$ one can rewrite this as

\[
u_t = \{ u , \partial_x^{-1} ~~ \}^{-1} u_x\,,
\]

\noindent so no $t$-derivatives of integrals appear on the right hand side.
This, however, is at the cost of introducing the inverse Poisson bracket
operator, and so
to avoid this the form (\ref{eq:sdein}) will be used (recall, if $\cal R$ is a
recursion operator then ${\cal R}^{-1}$ is also). Before proving
that $\{u,\partial^{-1}_t ~~\}$ is a recursion operator, the Lax pair
for the hierarchy (\ref{eq:sdeinhier}) will be found.

\section*{3. The Lax Equations for the anti-self-dual Einstein hierarchy}

One of the ways in which the construction of anti-self-dual Einstein metrics
may be understood is as follows \cite{Ashtekar.et.al,MasonNewman,Ward.c}.
Let ${\cal L}_0 = V_1 + \xi V_2$ and
${\cal L}_1 = V_3 + \xi V_4$ be commuting vector fields on some $4$-manifold
${\cal M}$ for all values
of the parameter $\xi\,$ (the spectral parameter), with each of the $V_i$
preserving a four-form $\omega$ on
${\cal M}\,.$ With $\Lambda=\omega(V_1,V_2,V_3,V_4)$ one may define a
(contravariant) metric

\[
g=\Lambda^{-1} (V_1 \phantom{_s}\otimes_s V_4-V_2 \phantom{_s}\otimes_s V_3
)\,,
\]

\noindent and this metric is automatically an anti-self-dual Einstein metric.
Conversely, all
such metrics can be constructed in this manner. In what follows it will be
convenient to defined a vector field $L_u\,,$ depending on some function
$u(t,x,y,z)$ by

\[
L_u = {\partial u\over\partial y}{\partial\phantom{z}\over\partial z} -
      {\partial u\over\partial z}{\partial\phantom{z}\over\partial y}\,.
\]

\noindent One can define the commutator of $L_u$ and $L_v$ in two different,
but equivalent ways:

\medskip

$\bullet$ Regard $L_u$ and $L_v$ as differential operators, and define
$[L_u,L_v]=L_u\, L_v - L_v\, L_u\,,$

$\bullet$ Regard $L_u$ and $L_v$ as vector fields, and define the
commutator to be the Lie bracket of vector fields.

\medskip

\noindent In each case one finds $[L_u,L_v]=L_{\{u,v\}}\,,$ where $\{u,v\}$
is the Poisson bracket defined by (\ref{eq:poisson}).

\medskip

What Grant \cite{Grant} did was to take

\begin{eqnarray*}
V_1 & = & {\phantom{-}}\partial_t\,, \\
V_2 & = & - L_u\,, \\
V_3 & = & {\phantom{-}} \partial_x\,, \\
V_4 & = & - \partial_t - L_v
\end{eqnarray*}

\medskip

\noindent and $\omega=dt\wedge dx\wedge dy\wedge dz\,.$ The condition that
the vector fields ${\cal L}_0$ and ${\cal L}_1$ commute results in
equations

\begin{eqnarray*}
u_t & = & \{ u , v \} \,, \\
v_t & = & u_x \,,
\end{eqnarray*}

\medskip

\noindent and on eliminating $v$ one obtains equation (\ref{eq:granteqn}).
An equivalent pair of vector fields, now
no longer linear in the constant $\xi$ are

\begin{eqnarray*}
{\cal L}_0 & = & \partial_t - \xi L_u\,, \\
{\cal L}_1 & = & \partial_x - \xi L_v - \xi^2 L_ u\,,
\end{eqnarray*}

\medskip

\noindent and for these to commute equation (\ref{eq:granteqn}) has to hold.
Motivated by the structure of the AKNS hierarchy, one can propose that the
anti-self-dual Einstein hierarchy be given by solving the
integrability conditions for the two vector fields

\begin{eqnarray*}
{\cal L}_0 & = & \partial_t - \xi L_u\,, \\
{\cal L}_1 & = & \partial_x - \sum_{r=1}^n \xi^r L_ {v_r}\,.
\end{eqnarray*}

\medskip The coefficient of the heighest power of $\xi$ in the commutator
$[{\cal L}_0,{\cal L}_1]$ is $L_{\{v_n,u\}}\,,$ and for this to be
zero one may take $v_n=u\,$ (in general, one has $v_n=\alpha u+\beta(t,x)$ but
this may be converted to the above form by various redefinition of the
variables). Using this result one finds, on equating the various
powers of $\xi\,,$ that

\begin{eqnarray*}
u_t & = & \{u,v_{n-1}\} \,,   \\
v_{i,t} & = & \{u,v_{i-1}\} \,,~~~~i=2\,,\ldots ,n-1\,, \\
v_{1,t} & = & u_x\,.
\end{eqnarray*}

\medskip

\noindent This system is also in Cauchy-Kovalevski form, and so formal
solutions
may be obtained as power series in $t\,.$
Systematically eliminating the $v_i$ in favour of the function
$u$ results in the single equation

\begin{eqnarray*}
u_t & = & \{ u,\partial^{-1}_t \{ u,\partial_t^{-1}\{\ldots\partial_t^{-1}
\{u,\partial_t^{-1},u_x\} \ldots \}\}\}\,, \\
& & \\
& = & \{ u, \partial_t^{-1} ~~\}^n u_x\,.
\end{eqnarray*}

\medskip

\noindent Thus the operator $\{u,\partial^{-1}_t ~~\}$ acts as a recursion
operator. In the next section the symmetry properties of this operator will
be considered.

\medskip

In the above one has a single scalar field $u(t,x,y,z)\,.$ One may extend
this to $m$-scalar fields by considering the vector fields

\begin{eqnarray*}
{\cal L}_0 & = & \partial_t - \sum_{r=1}^m \xi^r L_{u_r}\,, \\
{\cal L}_1 & = & \partial_t - \sum_{s=1}^n \xi^s L_{v_s}\,.
\end{eqnarray*}

\medskip

\noindent Without loss of generality one may take $n\geq m\,$ (otherwise just
interchange $x$ and $t$) and the integrability condition
$[{\cal L}_0,{\cal L}_1]=0$ can be solved at each power of $\xi\,,$ expressing
the $v_i$ in terms of the $u_i\,.$ This results in the system of equations
for the $m$-scalar fields

\medskip

\begin{equation}
\left( \begin{array}{c} u_1 \\ \vdots \\ u_m \end{array} \right)_t
= {\cal R}_m^{n-m}
\left( \begin{array}{c} u_1 \\ \vdots \\ u_m \end{array} \right)_x \,,
\label{eq:gensdhier}
\end{equation}

\medskip

\noindent where ${\cal R}_m$ is the matrix recursion operator

\medskip

\begin{equation}
{\cal R}_m = \left(
\begin{array}{ccccc}
\{u_{1{\phantom{-1}}},\partial_t^{-1}~~\}     &    1   & 0  & \ldots & 0
\\
\{u_{2{\phantom{-1}}},\partial_t^{-1}~~\}     &    0   & 1  & \ldots & 0
\\
\vdots                              & \vdots & \vdots  & \ddots & \vdots \\
\{u_{m-1},\partial_t^{-1}~~\} &    0   & 0       & \ldots & 1      \\
\{u_{m{\phantom{-1}}},\partial_t^{-1}~~\}     &    0   & 0  & \ldots & 0
\end{array}
\right)
\label{eq:genrecursionop}
\end{equation}

\medskip

\noindent Once again, the associated symmetry properties of this
generalised hierarchy will be considered in the next section.

\bigskip

\section*{4. Symmetries of the anti-self-dual Einstein hierarchy}

As mentioned in the introduction, one of the ways in which the recursion
operator may be introduced is via the study of the symmetries of the
original equation (see \cite{Magri} and \cite{FuchssteinerFokas}).
Given an evolution equation of the form

\[
u_t = K(u)
\]

\noindent a hidden symmetry $\tau$ corresponds to an infinitesimal parameter
satisfying

\begin{equation}
\tau_t = K'[\tau]\,,  \label{eq:symdef}
\end{equation}

\medskip

\noindent where $K'[\tau]$ is the Fr\'echet derivative of $K(u)\,.$ Given an
operator ${\cal R}$ satisfying the equation

\[
{\cal R}'[K] = [ K',{\cal R} ]
\]

\medskip

\noindent one may generate a hierarchy of symmetries, since

\begin{eqnarray*}
({\cal R}\tau)_t & = & {\cal R}_t \tau + {\cal R} \tau_t \\
             & = & {\cal R}'[K]\tau + {\cal R} K'[\tau] \\
             & = & K'[{\cal R}\tau]\,.
\end{eqnarray*}

\noindent Such an operator is said to be a strong symmetry, or recursion
operator. For ${\cal R}$ to be a strong symmetry for the hierarchy

\[
u_t = {\cal R}^n u_x
\]

\noindent it has to be a hereditary (or Nijenhuis) operator:

\[
({\cal R}'[{\cal R}f]g - {\cal R}{\cal R}'[f]g)-
({\cal R}'[{\cal R}g]f - {\cal R}{\cal R}'[g]f)=0\,.
\]

\medskip

\noindent In this section it will be shown that ${\cal R}_m\,,$ given by
(\ref{eq:genrecursionop}) is both a strong symmetry and a hereditary operator.
The proof will follow \cite{Cheng} where the symmetries of a
$2$-dimensional chiral model were studied.

\medskip

The field equations (\ref{eq:gensdhier}) are

\[
u_{i,t} = K({\bf u})_i
\]

\noindent where

\[
K({\bf u})_i = \left\{
\begin{array}{ll}
\{ u_i, \partial^{-1}_t u_{1,x} \} + u_{i+1,x} & i=1\,, \ldots\,,m-1\,, \\
& \\
\{ u_m, \partial^{-1}_t u_{1,x} \}             & i=m\,,
\end{array}
\right.
\]

\noindent and the conjectured recursion operator is

\[
({\cal R}{\bf f})_i = \left\{
\begin{array}{ll}
\{ u_i, \partial^{-1}_t f_1 \} + f_{i+1}\phantom{_x} & i=1\,, \ldots\,,m-1\,,
\\
& \\
\{ u_m, \partial^{-1}_t u_1 \}           & i=m\,.
\end{array}
\right.
\]

\medskip

\noindent The relevant Fr\'echet derivatives are

\[
K'[{\bf f}]_i = \left\{
\begin{array}{ll}
\{ f_i, \partial_t^{-1} u_{1,x}\}+\{ u_i, \partial_t^{-1} f_{1,x}\}+
f_{i+1,x} & i=1\,,\ldots\,,m-1\,, \\
& \\
\{ f_m, \partial_t^{-1} u_{1,x}\}+\{ u_n, \partial_t^{-1} f_{1,x}\}
& i=m\,,
\end{array}
\right.
\]

\medskip

\noindent and

\[
\begin{array}{ll}
({\cal R}[{\bf g}]{\bf f})_i=\{ g_i,\partial_t^{-1} f_1\} & i=1\,,\ldots\,,m\,.
\end{array}
\]

\medskip

\noindent These may be simplified by assuming $f_i=0$ and $u_i=0$ for
$i>m\,.$

\medskip

\indent $\bullet$ ${\cal R}$ is a strong symmetry.

\medskip

\noindent For arbitrary $m$-component vectors ${\bf f}$ and ${\bf g}$ one
has, by direct calculation,

\begin{eqnarray*}
({\cal R}'[K]{\bf f})_i & = & \{\{u_i,\partial_t^{-1} u_{1,x}\}+u_{i+1,x},
\partial_t^{-1} f_1\} \\
& & \\
({\cal R} K'[{\bf f}])_i & = & \{ u_i,\partial_t^{-1}(
\{f_1,\partial_t^{-1}u_{1,x}\}+\{u_1,\partial_t^{-1} f_{1,x}\}+f_{2,x})\}+\\
& & \{f_{i+1},\partial_t^{-1}u_{1,x}\}+\{u_{i+1},\partial_t^{-1}f_{1,x}\}+
f_{i+2,x} \\
& & \\
(K'[{\cal R}{\bf f}])_i & = & \{\{ u_i,\partial_t^{-1} f_1\} + f_{i+1},
\partial_t^{-1} u_{1,x}\} + \{u_i,\partial_t^{-1}
\{ u_1,\partial_t^{-1}f_1\}_x\} + \\
& & \{u_{i+1},\partial_t^{-1}f_1\}_x + \{u_i,\partial_t^{-1}f_2\}_x+f_{i+2,x}
\end{eqnarray*}

\medskip

\noindent By judicious use of the Jacobi identity, followed by an integration
by parts, it is straightforward to show that

\[
{\cal R}'[K]{\bf f} - K'[{\cal R}{\bf f}] + {\cal R} K'[{\bf f}] = 0 \,,
\]

\noindent and hence ${\cal R}$ is a strong symmetry, or a recursion operator.

\medskip

$\bullet$ ${\cal R}$ is an hereditary operator.

\medskip

\noindent Similarly

\begin{eqnarray*}
({\cal R}'[{\cal R}{\bf f}]{\bf g})_i & = & \{\{u_i,\partial^{-1}_t f_1\}+
f_{i+1},\partial_t^{-1} g_1\}\,, \\
&  & \\
({\cal R}{\cal R}'[{\bf f}]{\bf g})_i & = &
\{u_i,\partial_t^{-1}\{f_1,\partial_t^{-1} g_1\}\} +
\{f_{i+1},\partial_t^{-1}g_1\}\,,
\end{eqnarray*}

\medskip

\noindent and hence (again using the Jacobi identity followed by an integration
by parts)

\[
({\cal R}'[{\cal R}{\bf f}]{\bf g} - {\cal R}{\cal R}'[{\bf f}]{\bf g})-
({\cal R}'[{\cal R}{\bf g}]{\bf f} - {\cal R}{\cal R}'[{\bf g}]{\bf f})=0
\]

\noindent for arbitrary ${\bf f}$ and ${\bf g}$, thus showing ${\cal R}$
is a hereditary operator.

\medskip

Thus given any symmetry $\bf\tau$ one may construct a hierarchy of
related symmetries by the repeated application of ${\cal R}\,.$ It
thus remains to find some initial symmetries to start off this
procedure, and to find the Lie algebra generated by these symmetries.
For simplicity only the $m=1$ case will be considered, i.e. the anti-self-dual
Einstein hierarchy itself and not the more general system given by
(\ref{eq:gensdhier}).

\medskip

Two simple solutions of equation (\ref{eq:symdef}) are given by

\[
\begin{array}{rcl}
\tau_p^{(0)} & = & \{ u,p(y,z) \} \,, \\
\tau^{(0)}   & = & {\cal R} u
\end{array}
\]

\medskip

\noindent (note that although ${\cal R}u$ is a symmetry, $u$ is not, since
$2u_t=K'[u]$).
Thus, by the action of ${\cal R}\,,$ one has the following
hierarchies of symmetries:

\[
\left.
\begin{array}{rcl}
\tau_p^{(n)} & = & {\cal R}^n \tau_p^{(0)} \,, \\
\tau^{(n)}   & = & {\cal R}^n \tau^{(0)} \,.
\end{array}
\right\}
n = 0,1,2,\ldots .
\]

\medskip

\noindent Following the argument in \cite{Cheng} (which makes use of the
hereditary property of ${\cal R}\,$) one may easily show that the Lie
algebra of these symmetries is given by

\[
\left.
\begin{array}{rcl}
{[\![} \tau_p^{(m)} , \tau_q^{(n)} {]\!]} & = &-\tau_{\{p,q\}}^{(m+n+1)}\,, \\
{[\![} \tau_p^{(m)} , \tau^{(n)} {]\!]}   & = &\phantom{-}m\,\tau_{p}^{(m+n+1)}
\,, \\
{[\![} \tau^{(m)} , \tau^{(n)} {]\!]}     & = &\phantom{-} (m-n)\,\tau^{(m+n)}
\end{array}
\right\}
n=0,1,2,\ldots .
\]

\medskip

\noindent where $p$ and $q$ are functions of $y$ and $z$ alone, and the
Lie bracket $[\![F,G]\!]$ is defined by

\[
[\![ F,G ]\!] = F'[G] - G'[F]\,.
\]

\medskip

\noindent Similar results have been obtained by Grant \cite{Grant},
who has calculated the
Lie-point symmetries of equation (\ref{eq:granteqn}). The Lie point symmetries
of Pleba\~nski's equation (\ref{eq:plebanski}) has been calculated by
Boyer and Winternitz \cite{BoyerWinternitz}, and it would be interesting to
see how the Legendre
transformation that takes (\ref{eq:granteqn}) to (\ref{eq:plebanski})
connects the two sets of symmetries.

\medskip

Note that the recursion operator $\cal R$ factors,

\[
{\cal R} = {\cal H}_1 {\cal H}_0^{-1} \,,
\]

\noindent where

\begin{eqnarray*}
{\cal H}_0 & = & \partial_t \,, \\
{\cal H}_1 & = & \{ u, ~~ \} \,.
\end{eqnarray*}

\noindent This suggests a bi-Hamiltonian structure for the anti-self-dual
Einstein hierarchy. However one would need to study the corresponding
Hamiltonians in greater detail to verify this. The same splitting holds for the
more general hierarchy given by equations (\ref{eq:gensdhier}) and
(\ref{eq:genrecursionop}).

\bigskip

\section*{5. Comments}

In this paper the symmetry structure of the anti-self-dual Einstein equations
have been investigated. This work raises a number of further question, mainly
connected to the geometry of both the equations and its symmetries. The
dressing properties of the Lax pair for the self-dual Einstein equations
(in Pleba\~nski's form) and its reductions has been extensively studied by
Takasaki and Takebe \cite{TakasakiTakebe}, who has
also found the $\tau$-function for the system. It therefore seems important
to generalize these ideas to the hierarchy considered here, and to
understand the action of the symmetries in terms of the dressing operation.

\medskip

If the Poisson bracket $\{A,B\}$ were to be replaced by the matrix commutator
$[A,B]$ (with the now matrix valued fields being functions of $x$ and $t$
alone) then the results of this paper will still hold, the basic equations
being those of a Wess-Zumino-Witten topological field theory. Thus one
obtains a topological field theory hierarchy together with its symmetries.
In this case the geometry is much easier to understand, as the equations are
a reduction of the generalised self-dual Yang-Mills equations introduced in
\cite{Strachan}. Here the (flat) twistor space
${T}\cong{\cal O}(1)\oplus{\cal O}(1)$ (this being ${C}{P}^3$
with a single projective line removed) is replaced by
${T}_{m,n}\cong{\cal O}(m)\oplus{\cal O}(n)$ and the Ward construction
carries over naturally to this case. For the integrable systems under
consideration here what one needs is an analogous generalization of
Penrose's non-linear graviton construction \cite{Penrose}. One possibility
is to consider curved twistor spaces ${\cal T}_{m,n}$ with normal bundle
${\cal O}(m)\oplus{\cal O}(n)\,.$ Such spaces have been considered by
Gindikin \cite{Gindikin}, and it remains to see how the integrable
systems studied here fit into this framework.

\medskip

One may also construct a 3-dimensional hierarchy by assuming that the
K\"ahler potential has the form $\Omega=\Omega(\log w + \log x ,y,z)$
and a Legendre transformation takes (\ref{eq:plebanski}) to

\[
\nabla^2 u + {\partial^2 e^u\over\partial t^2} = 0
\]

\noindent (the so-called $SU(\infty)$-Toda, or Boyer-Finley equation) or to

\begin{equation}
h_{tt} = e^{-v} (h_{vs} h_{vt} - h_{vv} h_{st} )  \label{eq:boyerfinley}
\end{equation}

\noindent (where $v=\log w + \log x$). One may perform a similar reduction
to the hierarchy (\ref{eq:sdeinhier}), the lowest member of which is given
by (\ref{eq:boyerfinley}). It would be interesting to see if this hierarchy
corresponds to the $SU(\infty)$-Toda hierarchy constructed and studied by
Takasaki and Takebe \cite{TakasakiTakebe}.

\medskip

The symmetries of the full (i.e.\,with no restriction on the Weyl tensor)
Einstein equations have been studied recently by
Torre and Anderson \cite{TorreAnderson}, who
have found that no non-trivial generalised symmetries exist. Their approach
is coordinate independent, and it might be of interest to apply these
method to the anti-self-dual case.

\medskip

It is hoped that some of these issues will be addressed in a future paper.

\bigskip

\section*{Acknowledgements}

\medskip

I would like to thank James Grant for a number of useful conversations, and
also the University of Newcastle for a Wilfred Hall fellowship.

\end{document}